\documentclass{article}

\usepackage{microtype}
\usepackage{graphicx}
\usepackage{subfigure}
\usepackage{booktabs} % for professional tables

\usepackage{hyperref}

\usepackage[accepted]{icml2025}

\usepackage{amsmath}
\usepackage{amssymb}
\usepackage{mathtools}
\usepackage{amsthm}

\usepackage[capitalize,noabbrev]{cleveref}

\theoremstyle{plain}

\theoremstyle{definition}

\theoremstyle{remark}

\usepackage[textsize=tiny]{todonotes}

\icmltitlerunning{Brain-Inspired AI with Hyperbolic Geometry }

\begin{document}

\twocolumn[
\icmltitle{Brain-Inspired AI with Hyperbolic Geometry  }

\begin{icmlauthorlist}
\icmlauthor{Alexander Joseph}{sch}
\icmlauthor{Nathan Francis}{sch}
\icmlauthor{Meijke Balay}{sch}
\end{icmlauthorlist}
% It is OKAY to include author information, even for blind
% submissions: the style file will automatically remove it for you
% unless you've provided the [accepted] option to the icml2025
% package.

% List of affiliations: The first argument should be a (short)
% identifier you will use later to specify author affiliations
% Academic affiliations should list Department, University, City, Region, Country
% Industry affiliations should list Company, City, Region, Country

% You can specify symbols, otherwise they are numbered in order.
% Ideally, you should not use this facility. Affiliations will be numbered
% in order of appearance and this is the preferred way.
\icmlsetsymbol{equal}{*}

% You may provide any keywords that you
% find helpful for describing your paper; these are used to populate
% the "keywords" metadata in the PDF but will not be shown in the document
\icmlkeywords{Representation Learning, Brain-inspired AI, Neuroscience}

\vskip 0.3in
]

% this must go after the closing bracket ] following \twocolumn[ ...

% This command actually creates the footnote in the first column
% listing the affiliations and the copyright notice.
% The command takes one argument, which is text to display at the start of the footnote.
% The \icmlEqualContribution command is standard text for equal contribution.
% Remove it (just {}) if you do not need this facility.

% %\printAffiliationsAndNotice{}  % leave blank if no need to mention equal contribution
% \printAffiliationsAndNotice{\icmlEqualContribution} % otherwise use the standard text.

\begin{abstract}{Artificial neural networks (ANNs) were inspired by the architecture and functions of the human brain and have revolutionised the field of artificial intelligence (AI). Inspired by studies on the latent geometry of the brain, in this perspective paper we posit that an increase in the research and application of hyperbolic geometry in ANNs and machine learning will lead to increased accuracy, improved feature space representations and more efficient models across a range of tasks. We examine the structure and functions of the human brain, emphasising the correspondence between its scale-free hierarchical organization and hyperbolic geometry, and reflecting on the central role hyperbolic geometry plays in facilitating human intelligence. Empirical evidence indicates that hyperbolic neural networks outperform Euclidean models for tasks including natural language processing, computer vision and complex network analysis, requiring fewer parameters and exhibiting better generalisation. Despite its nascent adoption, hyperbolic geometry holds promise for improving machine learning models through brain-inspired geometric representations. 
}
\end{abstract}
\section{Introduction}
Artificial neural networks (ANNs), and more specifically, Deep Learning have led to great advancements in the field of AI in the last decade. The architecture of ANNs with connected neurons that fire based on activation mechanisms were inspired by the brain's architecture, albeit with many simplifications. This methodology has outperformed other AI approaches on a wide range of tasks and has led to deep learning systems powering many of the products we use in day-to-day life.\\

The vast majority of ANNs use Euclidean geometry as their latent geometry, with research on the use of other geometries being relatively niche, despite the latent geometry of various brain functions seemingly being hyperbolic. We propose that the latent geometric structures found in the brain can offer valuable insights for advancing AI models. This paper serves as a position paper, arguing for the increased integration of hyperbolic geometry in artificial neural networks; this argument is grounded in the idea that artificial intelligence can significantly benefit from adopting geometric representations that more closely resemble those used by the brain and seem to help it to efficiently and effectively perform cognitive tasks.

Hyperbolic geometry, a non-Euclidean geometry with negative curvature, is often referred to as the geometry of \textit{complex networks} \cite{Krioukov}. It has demonstrated significant advantages over other geometries in modeling complex networks and hierarchical structures. Complex networks are used to represent \textit{complex systems}—systems composed of many interacting parts whose collective behaviour cannot be trivially deduced from the behaviour of individual components \cite{Newman_2011}. These systems frequently exhibit hierarchical modularity \cite{siyari2018emergence} \cite{ravasz2003hierarchical} and are effectively modeled as complex networks. A prominent example of a hierarchical complex network is the human brain. 
\begin{table}[h!]
\centering
\begin{tabular}{ |p{2cm}||p{1.7cm}|p{1.7cm}|p{1.7cm}|  }
 \hline
 Property & Euclidean & Spherical & Hyperbolic\\
 \hline
 Curvature   & 0    & $> 0$ & $< 0$\\
  \hline
Parallel lines & 1    & 0 & $\infty$\\
 \hline 
 Triangles are & Normal    & Thick & Thin\\
 \hline
 Sum of $\triangle$ angles & $\pi$    & $> \pi$ & $< \pi$\\
 \hline
 Disk circumference & $2\pi r$    & $2\pi \sin \zeta r$ & $2\pi \sinh \zeta r$ \\
 \hline
 Disk area & $2\pi r^{2}/2$    & $2\pi (1 - \cos \zeta r)$ & $2\pi (1-\cosh\zeta r)$ \\
 \hline
\end{tabular}
\caption{Properties of Euclidean, spherical and hyperbolic geometries. $r$ is the radius of a Euclidean circle, and $\zeta = 1/\sqrt{|k|}$, where $k$ is the curvature.}
\label{table:1}
\end{table}

Recently, there has been an increase in research that has shown various brain functions as well as the brains architecture are best described with hyperbolic geometry \cite{sharpee_olfactory} \cite{zhang2023hippocampal} \cite{kim2023target}. The human brain is responsible for regulating bodily functions and enabling advanced cognitive processes such as reasoning and memory. Drawing inspiration from neuroscience has proven valuable in advancing AI systems \cite{ren2024brain}, and making use of the brains latent geometry could lead to further improvements.

%Recent findings suggest that hyperbolic rather than Euclidean geometry, better captures the brain's hierarchical and scale-free structure present in brain architecture and various cognitive functions.

Machine learning models which use hyperbolic geometry often demonstrate significant performance improvements over their Euclidean counterparts when applied to hierarchical data or scale-free data. Notably, these hyperbolic models frequently achieve comparable or superior results as well as requiring fewer parameters than Euclidean networks. This efficiency in parameter use, combined with enhanced performance, makes hyperbolic geometry an attractive approach in AI. However, there are barriers preventing researchers and data scientists from applying hyperbolic geometry in AI. We believe that research into hyperbolic geometry in AI should be encouraged, continuing to draw inspiration from human intelligence.\\

The rest of this paper is structured to provide an overview of hyperbolic geometry and its relation to both the brain and AI: The `Hyperbolic geometry' section provides a brief introduction to hyperbolic geometry, establishing the foundation for our discussion, the third section, `Hyperbolic geometry and the brain', explores the links between brain functions and hyperbolic geometry, `Hyperbolic Machine Learning' examines the current applications of hyperbolic geometry in AI, both for tasks that are analogous to those done with the human brain, as well as broader tasks. Finally, the `Future Work' section looks at various issues to overcome, and potential directions for future research and implementation.

\section{Hyperbolic geometry}
Geometry is one of the oldest branches of mathematics and it is concerned with the shape of individual objects, spatial relationships among various objects, and the properties of surrounding space.

In geometry, there are three types of spaces with constant curvature: spherical (positive curvature), Euclidean (zero curvature), and hyperbolic (negative curvature) (see Table \ref{table:1}). Euclidean geometry has been the dominant geometry in machine learning, mainly because its vector-space structure allows the application of powerful techniques of computational linear algebra. Another factor contributing to the widespread use of Euclidean geometry in machine learning may be familiarity bias, as curved geometries are less intuitive and harder to visualise. 

There are several models of hyperbolic geometry, each differing in their representation of lines (geodesics) and points of hyperbolic space. Common models are the Poincaré model, the Lorentz model (also known as hyperboloid), the Klein model and the upper half-space model \cite{cannon_1997}. In machine learning these models are applied based on their ease of implementation and the performance on a given task.

\begin{figure}[h!]
    \centering
    \includegraphics[width=0.45\textwidth]{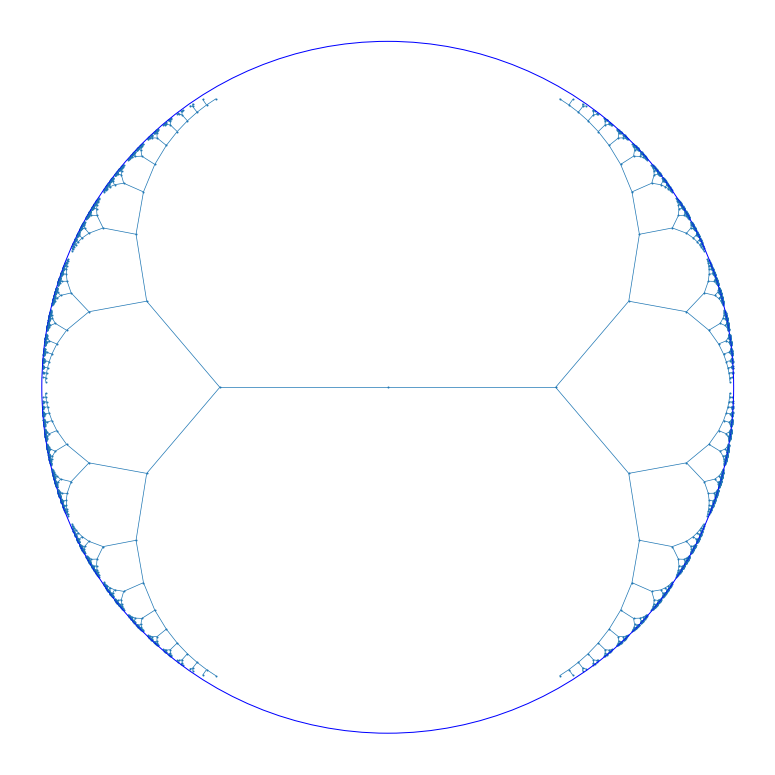}
    \caption{A binary tree embedded in 2D hyperbolic space. Large trees can be embedded with low distortion in hyperbolic space, but not in Euclidean space of the same dimension.}
    \label{fig:hyp-bin-tree}
\end{figure}

Two significant properties of hyperbolic geometry are its exponential expansion and its tree-like structure. This tree-like structure allows it to represent hierarchical structures with low distortion (see Figure \ref{fig:hyp-bin-tree}) and also model \textit{complex networks} very effectively, as complex networks often contain hierarchies \cite {ravasz2003hierarchical} \cite{Krioukov}.

The implication for machine learning – specifically representation learning – is that hyperbolic space naturally represents data with large-scale hierarchical structures. Hierarchical structures occur in numerous datasets, including those representing words, social networks, and biological neural networks. In particular, biological data is often represented using dendrograms or hierarchical tree structures \cite{sharpee_olfactory}. This widespread occurrence of hierarchical organisation in varied data types underscores the importance of methods capable of effectively handling such structures.

\section{Hyperbolic geometry and the brain}
Hyperbolic geometry has an important role in the physical architecture of the brain and the workings of various cognitive functions, both of which are integral to intelligence. This section first looks at the physical architecture of the brain, including the wiring of neurons and their interaction with each other. Secondly, we cover the latent geometry of cognition at both a macro and micro scale.

\subsection{Physical brain structure}
\label{physicalBrainStructure}

The physical architecture of the brain, which includes connections between neurons, affects information propagation through the brain, as well as the process of cognition.\\

The brain is a \textit{complex network} and displays traits typical of complex networks, such as being organised hierarchically. Hyperbolic geometry models these complex networks most closely which suggests the brain may be best described in hyperbolic space.
The maps of connections between neurons provide us with significant insight into the brain’s structure. To our knowledge, any attempts to show that the node connectivity of the human brain is Euclidean have failed \cite{cacciola2017coalescent}. When functions and structures of the brain are modelled using hyperbolic geometry instead of Euclidean geometry, this process has shown vast improvements, as shown in \cite{sharpee_olfactory} \cite{allard2020navigable}. This is consistent with the characterisation of hyperbolic geometry as the ``effective geometry of complex networks'' \cite{Krioukov}. 

The idea that complex systems have a hierarchical modular organisation originated in the early 1960s and results show that functional networks of the human brain have a hierarchical modular organisation \cite{meunier2009hierarchical}. Brain regions tend to coordinate by forming a highly hierarchical chain-like structure of homogeneously clustered anatomical areas \cite{mastrandrea2017organization}. The billions of neurons in the brain each resemble a tree with their structure, axons and dendrites \cite{Lewis_2016} and form a forest of trees as they connect to each other in the brain. As hyperbolic space is also known as a continuous version of trees, the tree-like hierarchical structure of the neurons could be best modelled in hyperbolic space.

The understanding of cellular differentiation (the transition of immature cells into specialised types) has also been helped with hyperbolic geometry; hyperbolic embeddings resulted in state-of-the-art representations of cell trajectories \cite{klimovskaia2020poincare}.
 
Connectomes are a comprehensive map of neural connections in the brain, and may be thought of as its “wiring diagram”. They were shown to be best modelled using hyperbolic geometry as they had a more accurate cartography of the brain than Euclidean geometry \cite{allard2020navigable}. \cite{cacciola2017coalescent} employed coalescent embeddings, a technique for mapping complex networks in the hyperbolic space, to reconstruct the latent geometry of the brain with a high degree of accuracy. This unsupervised approach not only provided insights into the brain's structural organisation but also demonstrated the ability to detect pathological alterations in the connectomes of individuals with Parkinson's disease. They noted that their results suggest that human structural brain networks are weakly hyperbolic.% 

\cite{sharpee2019argument} highlighted the significance of three-dimensional hyperbolic space in neural signalling and suggested that this geometric framework offers enhanced robustness compared to other dimensional representations, making it particularly suitable for describing the complex communication patterns among neurons. It was also demonstrated that neurons do not primarily interact with their Euclidean neighbours, but instead form what appears to be a relay network governed by non-Euclidean geometric principles. This behaviour is captured by hyperbolic models, which can generate near-perfect representations of neuronal connections as they exist in reality. In contrast, traditional Euclidean models often produce distorted and inaccurate maps of these neural networks. These findings underscore the potential of hyperbolic geometry to provide more faithful and insightful representations of the brain's intricate communication structure. \cite{whi2022hyperbolic} further proved that functional brain networks are best represented in the hyperbolic disc as they successfully embedded functional brain graphs observed on resting-state functional MRI (rs-fMRI) to a 2-dimensional hyperbolic plane (disc), with better fidelity compared with other product manifolds of low dimension. 

The connection between hyperbolic geometry and the physical can also be appropriately applied to other biological domains. Biological structures which aim to maximise their surface area within a given volume, such as trees and the gills in fish, are often highly crinkled, resembling embedded hyperbolic surfaces. This is particularly evident in the human brain, with its highly crinkled surface structure. As \cite{allard2020navigable} aptly stated, ``Hyperbolic geometry is a very natural way to represent the structural complexity of the brain''.

\subsection{Cognition}
\label{cognition}

Cognition can be defined as the states and processes involved in knowing, which in their completeness include perception and judgement. Cognition includes all conscious and unconscious processes by which knowledge is accumulated, such as perceiving, recognising, conceiving, and reasoning \cite{britannica_2024}. As the brain has a physically hierarchical structure, various cognitive functions of the brain may also be more accurately described in hyperbolic space as they rely on the interactions between neurons within its structure.

This section will consider the latent geometry of cognition from a comprehensive perspective presenting the literature that can provide insight into cognition at the macro scale of the brain. We will then consider major cognitive processes such as smell, vision, and language.

\subsubsection{Macroscopic Cognition}
\label{macroscopicCognition}

\begin{figure}
    \centering
        \includegraphics[width=1\linewidth]{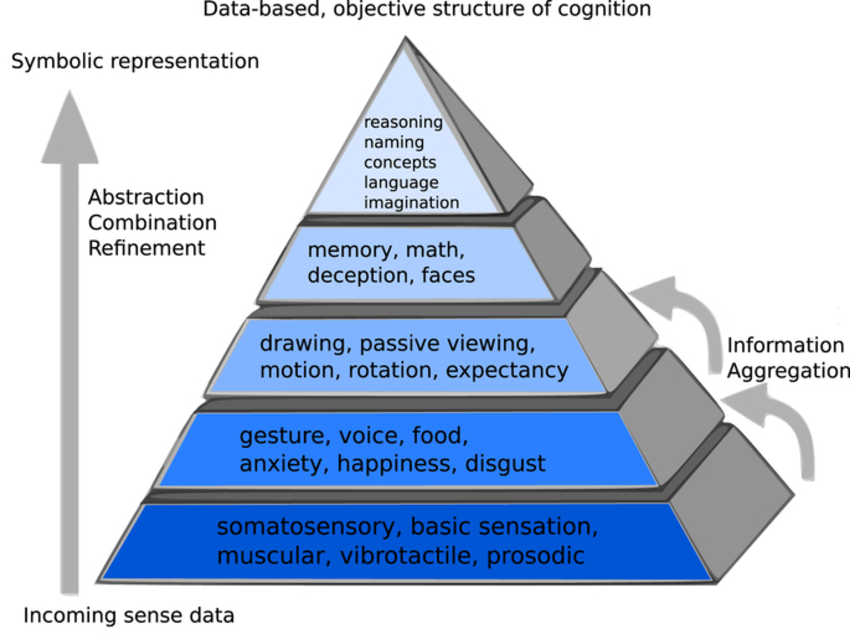}
    \caption{Oversimplified graphical model of the information representation flow from sensory inputs to abstract representations in human cortex \cite{taylor2015global}}
    \label{fig:cognition-levels}
\end{figure}

Mental states emerge from the interaction between physical and functional levels in the human brain. The human mind is what thinks, feels, perceives, imagines, remembers, and wills, encompassing our understanding of mental phenomena. It is a complex phenomenon built on the physical scaffolding of the brain \cite{bassett2011understanding}.\\
\cite{taylor2015global} revealed a hierarchical structure of cognitive processes in the brain (see Figure \ref{fig:cognition-levels}). Their research demonstrates that cognition operates through a series of levels. At the foundation lie primary sensory relay and processing functions. Moving upward, these sensory inputs are progressively integrated and abstracted. At the highest levels, we find complex cognitive functions such as reasoning and language. This hierarchical organisation suggests that higher-level cognitive processes emerge from the aggregation and abstraction of lower-level sensory information. Functions like language and reasoning reflect the brain's ability to construct complex understanding from basic inputs.

\subsubsection{Smell}
\label{smell}

In a study of human smell perception \cite{sharpee_olfactory}, it was demonstrated that smell perception is best modelled with hyperbolic geometry. The study showed that both natural odours and human perceptual descriptions of smells are best described using a three-dimensional hyperbolic space. This match in geometries can avoid distortions that would otherwise arise when using geometries ill-suited to mapping odours to perception. These findings suggest that the brain groups odours according to how often they occur together, rather than according to their molecular makeup. When a map of these odour clusters was made, it was found that the distance between similarly structured molecules was best represented using ideas of distance from hyperbolic geometry, rather than Euclidean. The study suggests hyperbolic perceptual organisation is likely general across sensory modalities. The link between smell perception and hyperbolic geometry arises from odour production in nature, where biochemical networks form tree-like structures that are well-represented in hyperbolic spaces.

\subsubsection{Spatial Representation}

Research into the brain's spatial encoding and navigation capabilities has been an active area of research for decades. 

Spatial representation, a fundamental cognitive function, refers to the brain’s ability to construct and utilise internal maps of the physical world. Research has focused on the CA1 region of the hippocampus, particularly in rats, due to its critical role in spatial memory and navigation. The similarity between a rat and a human brain has been evidenced by \cite{kim2023target} which suggests that the knowledge generated from decades of rodent research is relevant to human neurophysiology. Building on this foundation, recent research by \cite{zhang2023hippocampal} has unveiled new dimensions of complexity in spatial representation. It was discovered that neurons in the CA1 region of the rat hippocampus encode space through a non-linear hyperbolic geometry. This dynamic hyperbolic mapping allows neural circuits to efficiently adapt to changes in the environment and the rat’s familiarity with it, offering a sophisticated mechanism for spatial perception. 

The adoption of hyperbolic geometry in spatial navigation offers advantages that enhance both the efficiency and flexibility of neural computations. One of these advantages is its facilitation of hierarchical planning. This hierarchical organisation in neural networks enables a maximally informative representation of input signals allowing the brain to process and prioritise information in a way that supports efficient decision-making and problem-solving. Additionally, hyperbolic geometry supports efficient routing of signals, particularly in scenarios where network connections may change dynamically. This adaptability is crucial for navigating complex and ever-changing environments.

The dorsal CA1 region of the hippocampus is critical for spatial representation and analysis by \cite{zhang2023hippocampal} revealed that the size of the hyperbolic representation in this brain region increases as an animal becomes more familiar with its environment. This finding suggests that the brain’s spatial encoding mechanisms are not static, but instead dynamically expand with experience, thus enhancing the brain’s ability to navigate and understand increasingly complex spaces.

\subsubsection{Vision}
\label{vision}

The human visual system is a sophisticated network that allows for the perception and interpretation of the surrounding environment. At its core is binocular vision, which utilises the input from both eyes, enabling depth perception and a comprehensive field of view. Binocular vision depends on several factors, such as the motor system that coordinates eye movements \cite{EVANS20222}. Its primary advantage is to help provide a sense of distance and relative depth as we move around in our environment. As early as 1947, Luneburg developed a mathematical theory of binocular vision, concluding that the space of binocular vision has constant negative curvature, favouring hyperbolic geometry over Euclidean or elliptic geometries \cite{luneburg1947mathematical}. While his findings have since been contested, the visual space has been shown to have a hyperbolic geometry at least in part.

Studies have evidenced hyperbolic geometry and hierarchical structures in vision and visual perception. In \cite{Bilteanu2021} researchers posited a hyperbolic geometry framework for human vision, which demonstrated several important geometric features of ocular structures and visual functions. Using this framework the distribution of cone cells in the retina, responsible for colour vision, was modeled in hyperbolic space using a method that aligns well with microscopic studies of the retinal structure \cite{Bilteanu2021}. 

The visual cortex plays a critical role in visual processes. Located in the occipital lobe, the visual cortex is the primary cortical region responsible for receiving, integrating, and processing visual information from the retinas. It is divided into five distinct areas (V1 to V5), each with specialized functions and structures. The primary visual area (V1) is the first stage of cortical processing and contains a complete map of the visual field as seen by the eyes \cite{Huff2024}. \cite{Chossat2009} determined that the shape of area V1 is hyperbolic, describing it as a hyperbolic planform. They determined this by predicting brain activity patterns and designing experiments to test these hypotheses.

Given that hierarchies are best expressed in hyperbolic geometry, the link between hyperbolic geometry and the human visual system is further reinforced by the prevalence of hierarchical structures in visual motion perception. \cite{bill2020hierarchical}.

\subsubsection{Language}
\label{language}

Language serves as a cornerstone for human cognition \cite{berwick2013evolution} and evidence suggests that some elements of human language are best modelled using hyperbolic geometry. For example, the semantic relations between words in the hierarchical WordNet dataset were shown to be best modelled using hyperbolic geometry \cite{nickel2017poincare}, and research in the last two decades has suggested that hierarchical phrase structures characterise language. This is seen in the English language where noun and verb phrases are arranged within a clause in a hierarchical way rather than a linear order of words \cite{moro2000dynamic} \cite{friederici2011neural}.

Power laws, which are a functional relationship between two quantities, where a relative change in one quantity results in a relative change in the other quantity proportional to a power of the change, have been shown to appear in human language, and hyperbolic geometry can represent power laws well. An example of a power law in language is Zipf’s law, which states that the frequency at which a word appears in a given corpus is inversely proportional to its rank. Zipf’s law can be represented as a hierarchy with a cascade structure and has been noted as a signature of hyperbolic geometry \cite{chen2012zipf} \cite{sharpee2019argument}  \cite{mora2011biological} \cite{zipf1949principle}.

\paragraph{Other cognitive functions}

Studies have validated the claim that the brain also utilises hierarchical representations in other areas. Audio perception involves a complex, multi-stage process organised hierarchically in the human auditory cortex \cite{Okada2010}. This hierarchical structure is crucial for understanding how the brain processes and interprets sounds, ranging from simple tones to complex speech. There is further evidence in speech perception where the human brain continuously predicts a hierarchy of representations across multiple timescales \cite{Caucheteux2023}. 
 
\section{Hyperbolic Machine Learning}

The previous section suggests that many of the brain’s core cognitive functions — including vision, spatial navigation, and smell — may not only be conveniently modelled in hyperbolic space, but may be intrinsically encoded within it. For example, the organisation of olfactory perception might not only be better represented in hyperbolic space but could \textit{require} it, given the hierarchical, combinatorial nature of odours in the natural world.  Neural circuits operating in hyperbolic space offer distinct advantages, including maximal responsiveness to perturbations, allowing rapid adaptation to both external and internal changes \cite{sharpee2019argument}. This raises the possibility that hyperbolic space is a core geometric feature of brain processing, which could be instrumental in designing artificial neural networks that enhance accuracy, efficiency, and the ability to abstract information in ways that mirror human intelligence. 

As the field of AI has taken large inspiration from the human brain, we could use hyperbolic geometry as further inspiration. The use of hyperbolic geometry in machine learning and computer science, although limited, has been slowly growing in recent years. Hyperbolic machine learning models have demonstrated several advantages over their Euclidean counterparts: they require fewer parameters, can model tree-like structures without distortion, and more effectively fit complex, hierarchical data. The reduced number of parameters leads to shorter training times, providing an additional incentive for further research in this area. Moreover, deep learning models can struggle with capturing context and generalising to unseen data. Hyperbolic geometry, by better modelling the hierarchical and modular nature of many problems, may offer improved generalisation capabilities. This potential for enhanced performance and efficiency makes hyperbolic neural networks an increasingly important area of study in AI research.

The rest of this section will cover the existing literature in the field, the available tools practitioners can make use of and the issues practitioners currently face.

\subsection{Existing research}
\label{hyperbolic_ml_research}

Research showcasing the application of hyperbolic geometry on a range of tasks including computer vision, NLP, graph learning textual embeddings and symbolic learning have been undertaken in the field. These applications are particularly relevant for problems involving complex systems, hierarchical data structures, and cognitive functions typically associated with human intelligence.

In \cite{khrulkov2020hyperbolic}, image embeddings made using the Poincaré model of hyperbolic geometry have been shown to improve the performance of image classification, one-shot, and few-shot learning, as well as person re-identification, when compared to embeddings using Euclidean geometry. The effectiveness of hyperbolic geometry in this context is primarily attributed to its capacity to capture and represent the inherent hierarchical structures present in images. This hierarchical representation is advantageous when developing machine learning models for image analysis and understanding. The inspiration for the work came from the hierarchical relations that exist between images that are common in computer vision tasks. To build hyperbolic image embeddings \cite{khrulkov2020hyperbolic} used a neural network and changed the operations in the model to use operations in hyperbolic space. They also introduced an approach to evaluate the hyperbolicity of a dataset based on the concept of Gromov-hyperbolicity; this approach is useful for determining if hyperbolic geometry is appropriate for a dataset. Other recent approaches to finding whether hierarchies exist in a dataset include \cite{yang2024uhcone} which captures implicit hierarchical structures in data using hyperbolic embeddings and cone constraints. 

The use of a hyperbolic distance metric in computer vision tasks is effective in Open World Object Detection (OWOD). OWOD is the task of detecting both known and unknown objects in a scene while integrating learned knowledge for future tasks. In a novel method for OWOD, \cite{doan2024hyp} used a hierarchical approach to detect unknown objects in a scene. The hyperbolic distance metric was used because of its effectiveness in modelling hierarchies and this method showed up to a 6\% improvement compared to existing methods.\\

\begin{figure}
    \centering
    \includegraphics[width=1\linewidth]{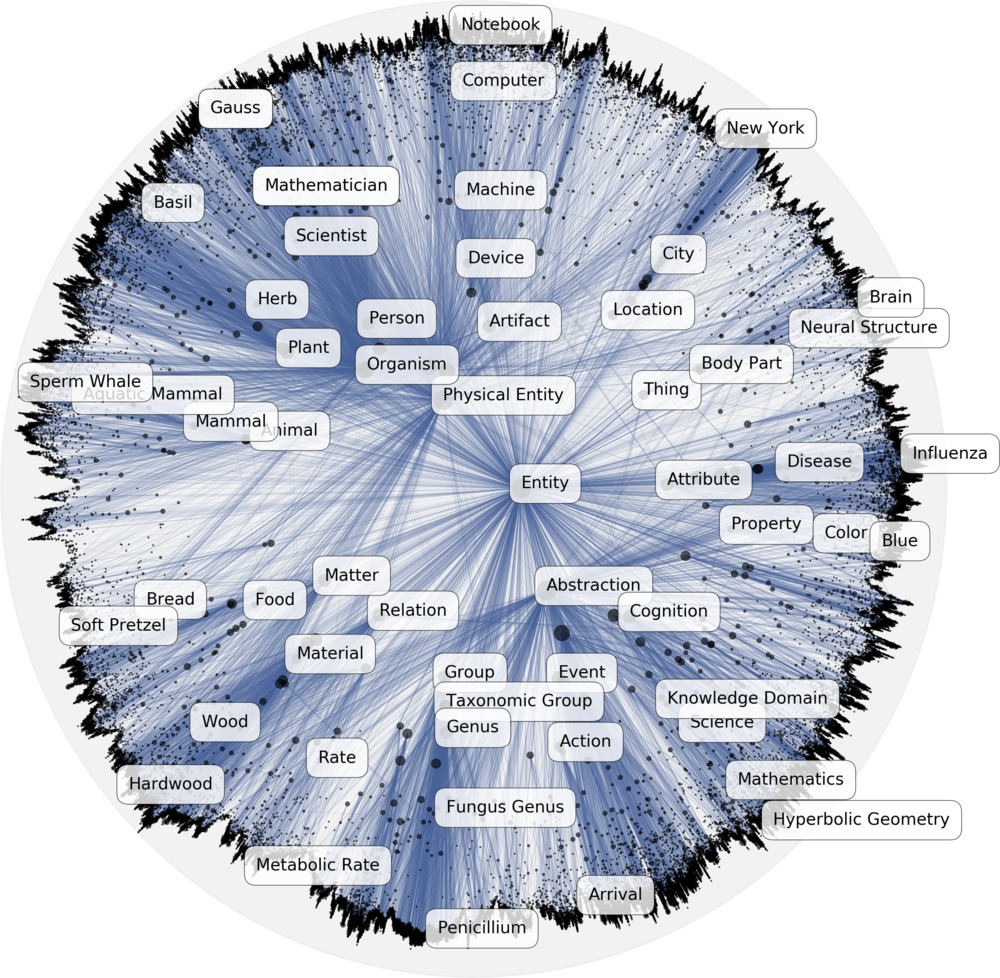}
    \caption{Wordnet nouns embedded in hyperbolic space \cite{poincare_embeddings}}
    \label{fig:wn-nouns}
\end{figure}

%\begin{sloppypar}
The ability to learn effective representations of symbolic data, including text, graphs, and multi-relational information, has become increasingly crucial in machine learning. These learned representations serve as foundational elements for various downstream tasks. Symbolic datasets that contain complex symbolic structures frequently exhibit latent hierarchies.\\
%\end{sloppypar}
\cite{nickel2017poincare} demonstrated the efficacy of hyperbolic embeddings for symbolic data, utilising the Poincaré model to achieve greater performance compared to its Euclidean counterparts. Their approach employed Riemannian optimization techniques, notably Riemannian Stochastic Gradient Descent (RSGD), to minimise the distance between symbolic representations. The experiments primarily made use of WordNet noun hierarchies, employing the embeddings for link prediction and reconstruction tasks. Poincaré embeddings (see Figure \ref{fig:wn-nouns}) demonstrated superior performance over their Euclidean counterparts, achieving comparable results with significantly fewer dimensions. Similar substantial improvements in performance were observed when applying these techniques to social network datasets for reconstruction and link prediction tasks. \cite{nickel2018learning} extended their previous work by using the Lorentz model. This approach led to an improvement on their earlier hyperbolic WordNet embeddings, which served to further demonstrate that hyperbolic embeddings outperform Euclidean embeddings. They also applied this approach to embedding cognates to find relationships between languages in modelling historical linguistics, and the hyperbolic embeddings outperformed the Euclidean counterpart once again.

Large Language Models (LLMs) have been at the forefront of a renewed interest in AI. Early research has looked at the effectiveness of LLMs which use hyperbolic geometry as their latent geometry. \cite{ChenHLHXZLS24} propose that pre-trained language models (PLMs) should operate entirely within hyperbolic spaces, as structured features within language are better encoded in hyperbolic than Euclidean spaces. Experiments demonstrated the superiority of hyperbolic PLMs over their Euclidean counterparts across a wide variety of tasks, indicating that the geometry of model representation is crucial for enhancement. \cite{he2024language} introduced Hierarchy Transformer encoders, which re-train transformer encoder-based language models to interpret and encode hierarchies using two innovations; hyperbolic clustering, and centripetal losses. Hyperbolic Centripetal Loss ensures parent entities are positioned closer to the Poincaré ball’s origin than their child counterparts. Hyperbolic Clustering Loss aims at clustering related entities and distancing unrelated ones. This approach led to significant improvements for Multi-hop Inference and Mixed-hop Prediction tasks on WordNet compared to Euclidean counterparts.\\

\cite{liu2019hyperbolic} performed molecular property prediction using a Hyperbolic Graph Convolutional Neural Network (HGCN) using the ZINC dataset. 
The model was developed by replacing Euclidean operations with hyperbolic ones, utilising an exponential map and incorporating both Lorentz and Poincaré models. Experimental results showed that these hyperbolic approaches consistently outperformed Euclidean graph convolutional networks, with the Lorentz model outperforming Poincaré’s across all experiments. Notably, these hyperbolic models also surpassed existing state-of-the-art deep learning molecular property prediction models developed by the wider community. The HGCN was further applied to a large-scale complex graph problem: predicting price fluctuations for the underlying asset of the Ethereum blockchain, where nodes represent addresses in the blockchain. This application showcased one of the key benefits of hyperbolic representations — the ability to inspect the hierarchy learned by the network.
Consistent with previous findings, the Lorentz model significantly outperformed the Poincaré model, while Euclidean representations yielded the poorest results. This pattern held across various experimental settings, reinforcing the advantages of hyperbolic geometry in capturing complex hierarchical structures in data.\\

Hyperbolic geometry has also been applied to reinforcement learning, showing considerable improvements over Euclidean geometry \cite{jacimovic2024reinforcement},  \cite{cetin2022hyperbolic}. This improvement is largely due to reinforcement learning naturally aligning with the hierarchical structure of Markov decision processes, which is a decision-making model often used in reinforcement learning, and is embedded in hyperbolic space.

Given the growing interest and complexity in hyperbolic machine learning, numerous comprehensive surveys have been published, offering broader perspectives on the application of hyperbolic geometry in machine learning. These reviews, including works by \cite{yang2022hyperbolic} \cite{yang2023hyperbolic} \cite{peng2021hyperbolic} \cite{mettes2024hyperbolic}, provide valuable insights into the current state and future directions of this area of research.

\subsection{Limitations and existing tools}
\label{hyperbolic_tools}

Currently, there are four models used to perform calculations in hyperbolic geometry, but their optimal applications in machine learning have not been thoroughly studied. Most research tends to default to the Poincaré model. As well as this hyperbolic geometry doesn't always lead to performance increases with hierarchical or scale-free data and we need to better understand how to utilise it or when other geometries may be better suited. Several other challenges also persist, including mathematical precision errors, the lack of linear algebra in hyperbolic geometry, and the absence of a natural mean. These issues complicate both theoretical work and practical implementation, contributing to the problem of vanishing and exploding gradients during neural network training. 

To effectively build models that leverage hyperbolic geometry at scale, tools are needed to abstract complex mathematical computations for machine learning practitioners. Several such tools exist, varying in maturity: Hyperlib \cite{hyperlib}, geoopt \cite{geoopt}, manifold.js \cite{juliamanifold}, and HypLL \cite{hypll}. These libraries enable the implementation of machine learning components in hyperbolic space, and their continued development and use are encouraged.

\section{Future Work}

There are several avenues for future research exploring the theory and application of hyperbolic geometry. Important areas which aim to bridge gaps in current knowledge, address existing challenges, and leverage the properties of hyperbolic geometry, are detailed in this section.

\textbf{Choice of hyperbolic model}: Four models of hyperbolic geometry are commonly used: including Poincaré, Lorentz, Klein and upper half-space model. However, a clear understanding on when each model is appropriate has not been developed. Future research should evaluate the strengths and weaknesses of each model, their use in machine learning problems and the problem cases they are best suited to.

\textbf{Maturation of tools: }Currently, a few tools and libraries are available that allow users to create and use hyperbolic machine learning models, such as geoopt \cite{geoopt}, manifold.js \cite{juliamanifold},  \cite{hyperlib} and HypLL \cite{hypll}. Most of these libraries implement multiple hyperbolic models, Riemannian gradient descent methods and the basic building blocks of a neural network such as linear layers and activation functions. However more complex components such as hyperbolic attention mechanisms are often found in repositories used by researchers that are not easy to disentangle from the rest of the code. The implementation of these components, as well as the implementation of popular and proven hyperbolic neural networks will help these libraries become more practical for practitioners.

\textbf{Precision Errors}: Mathematical precision errors can occur when implementing hyperbolic neural networks. This issue contributes to unstable training, the occurrence of NaN values in the network, vanishing gradients and exploding gradients. To mitigate this issue implementations of networks often have to clip values to prevent large values or use numerical types with a larger capacity. Further research on how to maintain mathematical precision is important so that future research can focus on the application of hyperbolic geometry rather than fixing mathematical issues in models.

\textbf{Hyperbolic LLMs and Knowledge Graphs}: Enhancing LLM performance has largely been driven by scaling parameters and data. A promising direction for improvement without further scaling is adopting a hyperbolic feature space, which can provide more efficient representations of linguistic structures like syntactic trees or semantic hierarchies. This could lead to better natural language understanding and generation. Future research should focus on optimizing LLMs in hyperbolic spaces to achieve scalable, high-performance models without excessive computational cost. There has been recent work exploring the use of hyperbolic geometry with LLMs such as \cite{mandica2024hyperboliclearningmultimodallarge}, \cite{yang2024hyperbolicfinetuninglargelanguage}, \cite{yang2024enhancing}.

Hyperbolic geometry has also proven useful in knowledge graph (KG) embeddings, particularly for hierarchical structures \cite{chami2020low}. Integrating KG with LLMs can improve context understanding and reduce issues such as hallucinations by providing accurate domain-specific knowledge. However, more research is needed on combining hyperbolic geometry with KGs to improve their integration and structure, ultimately enhancing AI models’ performance and interpretability \cite{kau2024combining}.

\textbf{Human-Centric Tasks and AI Alignment: }The alignment between hyperbolic geometry and the latent geometry of the brain makes it a good choice for tasks related to human cognition and perception, including but not limited to modelling language, and predicting visual and sensory preferences. The use of hyperbolic geometry in modelling the brain is established in neuroscience. Still, in the field of AI this could increase, particularly given the prominence of tasks in AI that relate to the human brain. Moreover, designing ANNs that further reflect the brain's structure and functions could make them more aligned with other aspects of human intelligence.

Modelling human language and understanding sensory perceptions—such as determining which smells people might find appealing—involves modelling how the brain represents and processes information. Performance could improve by mimicking the brain’s hyperbolic cognitive representations more closely..

Works such as \cite{nonaka2021brain} have developed quantitative measures that score the alignment between deep natural network representation spaces and those in the brain for object recognition. This work could be extended beyond object recognition and applied to other human-centric tasks.

\section{Alternative Views}
\label{alternativeviews}

In this section we will address two key alternative views to those presented in this paper. 

\textbf{Alternative view 1: Both spherical and Euclidean geometry have roles that have been observed in the brain, and have shown to work well in AI systems. Hyperbolic geometry does not need to be advocated for in particular, apart from the fairly niche use cases it suits.}

Firstly, only hyperbolic geometry can be said to be broadly applicable to neural circuits in the brain \cite{sharpee2019argument}.

Secondly, each type of geometry has its advantages and is suitable for different tasks and types of data.

Spherical geometry is suited to data that is observing things over angles or directions,  such as data for autonomous vehicles, drones and also in the brain when looking at signals over directions. Euclidean space can be thought of as a linear space where the shortest distance between two points is a straight line. This often works well when the relationships between data points can be described by a straight line or is is flat. However, its effectiveness can lessen when this is not the case, and  with high-dimensional data where points in space are often not related by a straight line.
 
Hyperbolic geometry is suited to tree-like and hierarchical data, and these often appear in complex networks and systems. Humans exist within and navigate through a multitude of complex systems, from the ecosystems they inhabit to the economy, the internet, and social networks. These systems are not linear or simple; rather, they are hierarchical, multi-scaled, and densely interconnected. Euclidean and Spherical geometry, while useful for data they are suited to are insufficient for capturing the rich, intricate relationships and growth patterns that characterise complex systems. Hyperbolic geometry provides a more effective framework for modelling these systems, yet it does not have widespread use for these types of tasks. In cognitive tasks that involve layers of abstraction, generalisation, and the integration of multiple layers of meaning, hyperbolic representations offer a more compact and accurate way to capture these relationships.  

In building intelligent systems, decision-making is often modelled using Markov Decision Processes (MDPs), which naturally form hierarchical structures like decision trees. Hyperbolic geometry efficiently represents these hierarchical relationships, making it well-suited for MDPs.  This has been validated in recent reinforcement learning work where hyperbolic space has offered advantages over Euclidean space by enabling more efficient exploration and decision-making \cite{jacimovic2024reinforcement},  \cite{cetin2022hyperbolic}. 

For these reasons hyperbolic geometry enables a more comprehensive and flexible understanding of the world around us, making it crucial for building intelligent systems.  

% This is in note form
\textbf{Alternative view 2: Euclidean models have been the primary geometry used in machine learning to date and have achieved great success, particularly with scaling. Since Euclidean geometry seems to scales well there is no need to use move to hyperbolic geometry which is not trivial to implement and has a steep learning curve.:} 

Although Euclidean geometry doesn't fit all data types perfectly, it often works well when scaled, as seen with LLMs. Despite language’s hierarchical structure and improvements shown by hyperbolic LLMs, most models still rely on Euclidean space and perform well at larger sizes.

Implementing machine learning models with hyperbolic geometry is not trivial as it requires an understanding of geometry and writing calculations in hyperbolic space, and this is compounded by the fact that resources for AI practitioners are dominated by Euclidean geometry. Due to these hurdles, the advantages of hyperbolic geometry may not be viewed as worth the effort. While this will be true in some cases, and a better understanding of how much hyperbolic geometry may help any given task, hyperbolic geometry often requires fewer parameters, which leads to smaller models and shorter training time which will help to democratise AI innovations away from the big tech monopoly we have been heading towards. In addition to this, fields such as Topological Deep Learning and Geometric Deep Learning have shown the importance of making sure that a model's architecture aligns with the topology and geometry of the data it is dealing with.

\section{Conclusion}

With their architectures and functionalities initially inspired by the brain’s neuronal networks and activation mechanisms, ANNs have profoundly transformed the landscape of artificial intelligence. However, despite this biological inspiration, the latent geometry of neural networks has primarily been Euclidean, contrasting with the brain’s. Recent findings suggest that the human brain, with its hierarchical organisation and complex network properties, is best represented using hyperbolic geometry. The brain's inherent flexibility and efficiency is enabled by hyperbolic space, suggesting that ANN's could benefit from adopting similar representations. This led us to explore the use of hyperbolic geometry in AI systems for modelling datasets that contain hierarchies and complex networks. Due to the improvements seen in AI systems using hyperbolic geometry and the role of hyperbolic geometry in the brain, we propose that leveraging hyperbolic geometry could enable AI models to process information in a way that more closely mirrors human cognitive processes. 

Hyperbolic geometry’s exponential expansion of space and the characteristic thinness of triangles enable more efficient and less distorted representations of hierarchical data, making it particularly well-suited for modelling complex networks prevalent in biological systems, social networks, and linguistic structures. While traditional Euclidean neural networks have been highly successful, they face occasional limitations when it comes to efficiently capturing hierarchical structures. By integrating hyperbolic geometry into neural networks, research has demonstrated notable improvements across a variety of tasks, including image classification and molecular property prediction.

While promising, the adoption of hyperbolic geometry in AI is still in its early stages. Challenges like developing better tools, overcoming precision errors, and finding optimal applications remain. However, we hope this paper inspires further research into its potential to create more brain-like, efficient, and scalable systems. 

\section{Impact Statement}

This research proposes a fundamental shift in AI by aligning architectures with the brain's inherent geometric structure. By establishing hyperbolic geometry as central to human cognition, it suggests current AI may be limited by its reliance on Euclidean geometry. Incorporating hyperbolic geometry could lead to AI systems that better capture hierarchical, complex relationships, improving performance with fewer parameters. This approach offers a path toward more efficient, brain-like AI, potentially requiring fewer computational resources, while bridging neuroscience and AI to develop more advanced, interpretable systems aligned with human cognition.

\bibliographystyle{icml2025}
\bibliography{references} % Entries are in the refs.bib file

\end{document}